\documentclass[runningheads]{llncs}
\usepackage[T1]{fontenc}
\usepackage{hyperref}
\usepackage{graphicx}
\usepackage{subfigure}
\usepackage{subcaption}

\usepackage{graphicx}
\begin{document}
\title{Queuing for Civility: Regulating Emotions and Reducing Toxicity in Digital Discourse}
%
%
\author{Akriti Verma\inst{1} \and
Shama Islam\inst{1} \and
Valeh Moghaddam\inst{2}\and
Adnan Anwar\inst{2}}
\authorrunning{A. Verma et al.}
%
\institute{School of Engineering, Deakin University, Australia \and
School of Information Technology, Deakin University, Australia
}
\maketitle              
\begin{abstract}
The pervasiveness of online toxicity, including hate speech and trolling, disrupts digital interactions and online well-being. Previous research has mainly focused on post-hoc moderation, overlooking the real-time emotional dynamics of online conversations and the impact of users' emotions on others. This paper presents a graph-based framework to identify the need for emotion regulation within online conversations. This framework promotes self-reflection to manage emotional responses and encourage responsible behaviour in real time. Additionally, a comment queuing mechanism is proposed to address intentional trolls who exploit emotions to inflame conversations. This mechanism introduces a delay in publishing comments, giving users time to self-regulate before further engaging in the conversation and helping maintain emotional balance. Analysis of social media data from Twitter and Reddit demonstrates that the graph-based framework reduced toxicity by 12\%, while the comment queuing mechanism decreased the spread of anger by 15\%, with only 4\% of comments being temporarily held on average. These findings indicate that combining real-time emotion regulation with delayed moderation can significantly improve well-being in online environments.

\end{abstract}
\section{Introduction}
The fast pace of online interactions often leads to emotionally charged conversations, increasing the potential for polarised reactions and digital conflict \cite{smith2022digital}. The rising concern of online toxicity, which often includes hate speech and trolling, has negative effects on users’ well-being and the overall health of digital communities \cite{maarouf2022virality}, \cite{goel2016structural}. Current methods for managing online toxicity primarily focus on post-hoc moderation, where harmful content is identified and removed after it's been posted \cite{chandrasekharan2022quarantined}. Despite advancements in content moderation tools, there is still a need for proactive solutions that empower users to manage their emotional impact on conversations before toxicity escalates \cite{trujillo2022make}, \cite{gongane2022detection}.

Digital Emotion Regulation (DER), the use of digital technologies to influence one’s emotional state, has recently emerged as a habit individuals acknowledge and employ in everyday life by combining a variety of applications and devices for purposefully managing emotions \cite{wadley2020digital}, \cite{smith2022digital}. Some examples include listening to uplifting music while exercising, watching comedy or light-hearted videos to relieve stress after work, playing social video games when lonely, or scrolling through social media applications to combat boredom. While DER applications are widely acknowledged in personal contexts, they are underexplored in interactive online environments where emotions can propagate quickly and influence others \cite{manikonda2018twitter}. Additionally, existing solutions inadequately address the gap between reactive moderation and proactive emotional regulation. This paper introduces a new approach to promoting healthier online conversations by advocating self-reflection-based emotion regulation and integrating it with a comment queuing mechanism to curb the propagation of negative emotions, particularly in conversations vulnerable to trolling. 

The practice of self-reflection, which involves assessing one's own emotions and behaviour, has been widely acknowledged as a method for managing emotional well-being in everyday interactions \cite{herwig2010self}. However, its potential for regulating emotions in online settings has not been thoroughly explored. By encouraging self-reflection in digital spaces, users can gain greater awareness of how their comments affect conversations emotionally. This will allow them to reevaluate and manage their emotions before engaging further \cite{kiskola2021applying}. This proactive approach shifts the focus from reactive moderation, which deals with toxicity only after it has occurred, to real-time emotion regulation, thereby reducing the chances of emotional escalation.

This paper introduces a method that utilises a graph-based framework to visualise the emotional tone of a conversation and informs users of their impact on its emotional state. This approach promotes self-reflection, providing users with an opportunity to consider the emotional impact of their comments. Additionally, in cases of deliberate trolling, a comment queuing mechanism is proposed. This system introduces a brief delay in publishing comments, allowing users to reflect on their emotional state during the pause, while also ensuring the emotional balance of the conversation is maintained.

Our analysis of social media data shows that promoting online self-reflection can be a powerful tool for enhancing emotion regulation, especially in navigating emotionally charged conversations. This paper highlights the value of Implicit Emotion Regulation through Self-Reflection and deliberately delayed responses by offering an opportunity for cognitive reappraisal \cite{kiskola2022online}. This subtly nudges users toward a more considerate and regulated response, aiming to reduce online toxicity and facilitate healthier digital spaces. Therefore, this work makes the following research contributions: 
\begin{itemize}
    \item A novel self-reflection-based DER system: This paper proposes a graph-based framework that informs users of their emotional influence on a conversation and promotes implicit emotion regulation through a user-centred approach.
    \item A comment queuing mechanism for real-time emotion regulation: This paper introduces a comment queuing system that delays user responses to prevent impulsive emotional reactions, offering users time to reflect on their emotional state. Unlike existing delay mechanisms, our system is adaptive and context-sensitive, dynamically adjusting based on the emotional dynamics of the conversation.
    \item Empirical validation of emotion regulation in digital conversations: Our initial findings show a 12\% reduction in the spread of hate speech and anger, exceeding Google’s Perspective API by 3\%. Furthermore, when using the queue mechanism during ongoing conversations, there is a potential decrease of up to 15\% in trolling and hate speech propagation, with only 4\% of comments temporarily held for an average of 47 seconds.
    
\end{itemize}

\section{Literature Review}
As social media platforms grow, there is an increasing need to address emotional well-being in online environments. This brief literature review explores recent research on DER, self-reflection as a tool for emotion regulation, trolling behaviour, and strategies for comment moderation. It highlights the significance of this research in advancing our understanding of these areas.
\subsection{Digital Emotion Regulation (DER)}
DER is a growing field of study within psychology and human-computer interaction, focusing on the influence of digital tools on individuals' emotional well-being. DER research examines the impact of digital technologies on emotion regulation (ER) \cite{mcrae2020emotion}, \cite{gross2015emotion}, \cite{gross2014emotion}, \cite{wadley2020digital}. Studies have utilized methods such as self-reports and diaries to demonstrate the use of digital technologies for everyday ER \cite{smith2022digital}, \cite{shen2020video}, \cite{hossain2022motivational}, \cite{lukoff2018makes}. DER interventions have incorporated biofeedback and reminder-based systems to enhance ER skills by encouraging users to practice ER \cite{kiskola2021applying}, \cite{kou2020emotion}. Additionally, multi-modal sensors, such as cameras and touch sensors, have been employed to observe and recognize emotional changes during the DER process \cite{yang2021behavioral}, \cite{ruensuk2020you}. Research has also revealed the concentration of toxic discussions particularly among individuals with limited social connections \cite{thomas2022s}, \cite{saveski2021structure}. Furthermore, studies have shown that moral emotions expressed in tweets can influence the circulation of false rumours, with key users playing a crucial role in initiating online social movements \cite{solovev2022moral}. However, much of the existing literature focuses on individual emotional regulation and overlooks the social dynamics of online conversations, where emotional contagion from others can significantly impact one's emotional state. This study addresses the gap in current DER literature by introducing an approach that includes emotion regulation strategies and feedback mechanisms during online interactions.

\subsection{Emotion Regulation and Self-Reflection}
Emotion regulation, as defined by Gross \cite{gross2008emotion}, encompasses the processes through which individuals manage the emotions they experience, how they express them, and the timing of these emotions. Traditionally studied in offline environments, emotion regulation enhances interpersonal communication by enabling individuals to reconsider their emotional responses before reacting. Self-reflection is a crucial component of emotion regulation that empowers individuals to assess their emotions and underlying motivations, potentially leading to cognitive reappraisal \cite{ochsner2005cognitive}. Studies have found that temporal distancing—taking time before responding to emotional stimuli—can decrease emotional arousal and promote better emotional control \cite{grossmann2014exploring}. While this concept has mainly been explored in face-to-face and other offline settings, it is seldom put into practice in digital interactions where the rapid pace of interaction often allows little time for reflection. Existing research has examined emotion regulation in digital contexts but primarily within educational tools or therapeutic applications, with minimal attention to public online discourse \cite{hadwin2017self}, \cite{hadwin2019academic}. Our work bridges this gap by developing a framework that promotes self-reflection in real-time online conversations, enabling users to pause, reflect, and regulate their emotions before posting a potentially toxic comment.

\subsection{Online Toxicity and Deliberate Trolling}
While labelling emotions may help control unintentional emotional outbursts, dealing with deliberate trolling is more complex \cite{jane2020online}. Trolling, which involves intentionally disrupting online conversations to provoke emotional reactions, has been extensively studied in social and computer science \cite{buckels2014trolls}. Studies have identified trolls as individuals who derive pleasure from causing emotional chaos and escalating conflict for a range of motivations from seeking amusement and attention to more malicious intentions such as causing harm or manipulating discussions \cite{kumar2017antisocial}, \cite{cheng2017anyone}. Current moderation approaches primarily focus on removing harmful content after it is posted, which is reactive rather than proactive. While tools such as banning a user and deletion of comments can mitigate some of the damage caused by trolls, they do not address the initial escalation \cite{yin2009detection}, \cite{chandrasekharan2017you}, \cite{chandrasekharan2022quarantined}. Prior research suggests that traditional content moderation often exacerbates trolling behaviour, as trolls enjoy pushing the boundaries of platform policies \cite{phillips2015we}. This highlights the need for a more preventive approach to managing trolls, which our work aims to achieve through a comment queuing mechanism designed to allow for self-reflection-based emotion regulation.

\subsection{Conversation analysis and comment moderation mechanisms}
Comment moderation has been the primary defence against online toxicity, typically employing the detection and removal of harmful content, as well as flagging inappropriate behaviour. Platforms such as Facebook and Twitter have implemented algorithmic tools, such as content filters and moderation bots, to mitigate toxic behaviour \cite{gorwa2020algorithmic}, \cite{gillespie2020content}. However, recent literature has highlighted their limited capacity to effect meaningful, long-term changes in user conduct \cite{gillespie2018custodians}. It has been argued that the root of the problem lies not just in identifying harmful content but in changing the emotional responses and behaviours that drive users to post it in the first place \cite{frischlich2021roots}, \cite{park2022measuring}.

In digital communication, effective Emotion Regulation (ER) includes the utilisation of factual cues, automatic emotion identification, and didactic learning \cite{kiskola2021applying}, \cite{slovak2022designing}. Implicit emotion regulation, which operates automatically and can be automated, has garnered recent attention. Affect labelling, making emotionally charged aspects of conversations more apparent, can enhance emotion regulation \cite{torre2018putting}. Current research has explored approaches to comment moderation, such as the introduction of delays aimed at limiting impulsive reactions. For example a delayed feedback loop can provide users time for reconsideration before posting their comments \cite{cheng2017anyone}. This aligns with emotion regulation theory, which argues that the introduction of time-based pauses can facilitate users in engaging in cognitive reappraisal, thereby defusing emotionally charged situations. Yet, most existing comment delay mechanisms are either arbitrarily timed or overlook the emotional dynamics in the conversation.

Therefore, the following gaps exist in the literature:
\begin{itemize}
\item Lack of Real-Time Emotion Regulation in Online Conversations: Current research explores Digital Emotion Regulation (DER) in personal use of digital tools, but lacks real-time strategies for regulating emotions in interactive online settings, focusing mainly on post-hoc content moderation rather than preventing emotional escalation during conversations.
\item Limited Application of Self-Reflection in Digital Spaces: Although self-reflection is commonly used for offline emotion regulation, its application to online communication is not well explored. Methods to integrate self-reflection into digital interactions can promote more thoughtful, emotionally regulated responses.
\item Limited Focus on the Root Causes of Trolling: Current research focuses on removing or moderating toxic content rather than addressing the root causes of trolling behaviour, such as emotional arousal and impulsivity. Strategies are needed to encourage users to pause and reflect before posting.
\end{itemize}

Our research extends upon these findings by integrating a comment queuing system with a graph-based framework designed to monitor the emotional tone of the conversation continually. This approach not only introduces temporal delays to encourage self-reflection but also ensures that the emotional equilibrium of the conversation is maintained, thereby reducing the probability of escalation resulting from trolling or impulsive behaviour.

\section{Methodology}
\begin{figure}[h]
  
    \centering
    \includegraphics[width=10.8cm,height=11cm,keepaspectratio]{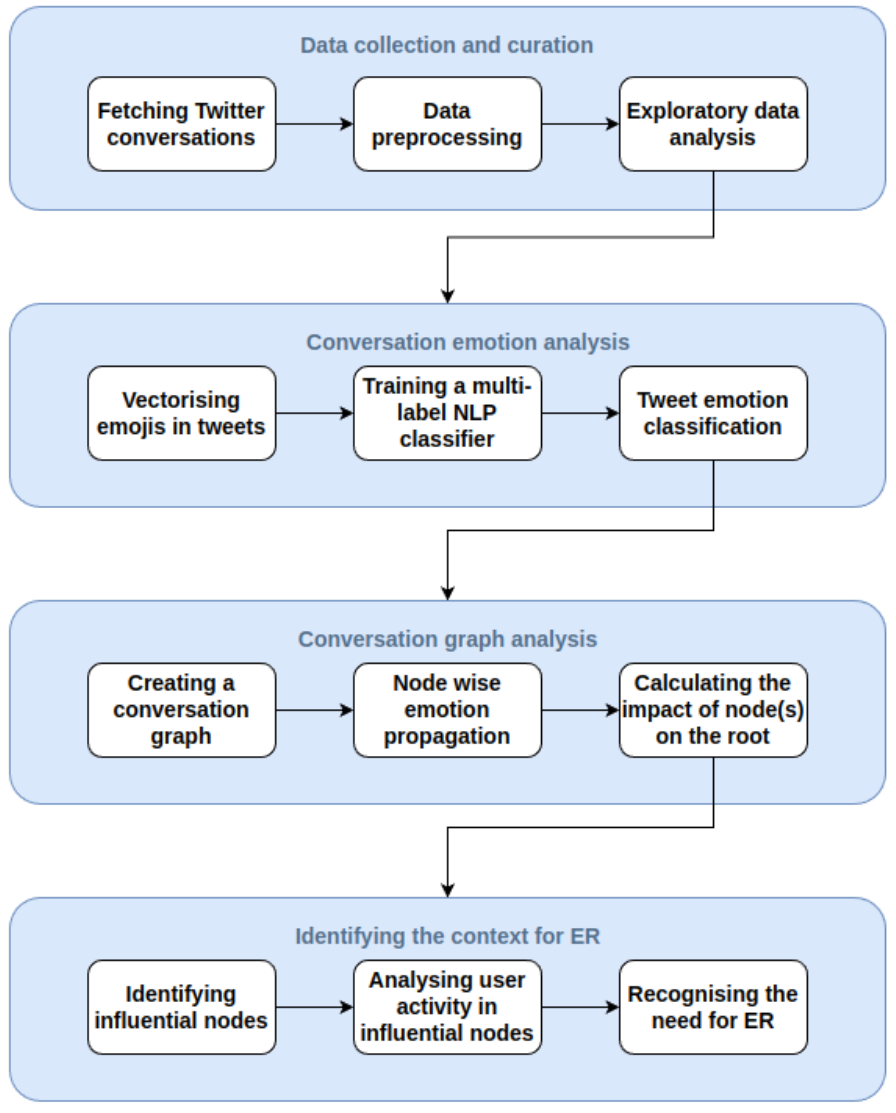}
  \caption{eImpact: Framework for supporting emotion regulation in social media conversations}
  \label{fig:Framework}
  \end{figure}

This paper presents an approach to managing emotions in online conversations through self-reflection. The methodology comprises four stages of the eImpact framework (Fig. \ref{fig:Framework}), including emotion detection, graph-based conversation analysis, and the implementation of a comment queuing system to encourage self-reflection and minimise emotional escalation. 

\subsection{Data Collection and Preprocessing}
For this study, we focused on data from Twitter and Reddit due to their distinct communication styles and ability to capture a wide range of emotional expressions.

Twitter facilitates rapid, public interactions that often lead to direct and emotionally charged conversations, particularly around political and social issues. Its fast-paced nature makes it well-suited for analysing how emotions propagate and the need for self-reflection in digital interactions \cite{manikonda2018twitter}.

On the other hand, Reddit, with its longer posts and topic-specific communities, allows for more in-depth discussions. Its threaded structure enables the tracking of emotional development across conversations, providing richer data for the analysis of emotion regulation \cite{manikonda2018twitter}.

\subsubsection{Twitter Dataset} Data was collected from Twitter conversations involving tweets from members of the Australian Parliament between April 2020 and August 2022 using the  \href{https://developer.twitter.com/en/docs/twitter-api}{Twitter API (Tweepy)}. These tweets, which contain discussions on COVID-19, policy changes, and other political topics, were gathered and refined for analysis. The dataset comprises 25K tweets, providing insights into public reactions and interactions expressing various emotions.

\subsubsection{Reddit Dataset} For more in-depth conversations, we acquired 15,000 user posts and comments from Reddit using the \href{https://praw.readthedocs.io/en/stable/}{Reddit API (PRAW)}. The dataset contains posts, replies, comments, and timestamps, allowing for the tracking of emotional shifts throughout the conversations. All text was cleaned to eliminate extraneous characters, links, and symbols, and was tokenized for natural language processing (NLP) analysis.

\subsection{Emotion Classification}
To identify emotions in the collected data, we utilised the NRC Word-Emotion Association Lexicon \cite{mohammad2013nrc}, which recognises emotions such as anger, fear, anticipation, trust, surprise, sadness, joy, and disgust. The classification process involved analysing text and emojis by converting them into vectors. The emojis in the tweets were substituted with vector representations created by \href{https://radimrehurek.com/gensim/}{Gensim} through the Emojinal library \cite{barry2021emojional}, after which the tweet text was tokenized using the \href{https://www.nltk.org/api/nltk.tokenize.casual.html}{TweetTokenizer}. Each comment received an emotion and intensity score ranging from 0.1 to 1.0, reflecting the emotion and strength of the expressed emotion in its text.

\subsection{Graph-Based Conversation Analysis}
Conversations were represented using a directed acyclic graph (DAG) structure, where original posts function as root nodes, while replies and comments act as child nodes. The connections between comments and replies are represented by edges. This hierarchical structure is then used to assess the influence of comments on the overall conversation.

The eImpact model represents every comment or tweet as a node in the graph, with each node being allocated an emotion score \cite{10520471}. This score is determined by the likelihood that the emotion classification model associates with the content. The influence of each node on the root node (original post) is evaluated using various metrics:

\begin{itemize}
\item Number of Replies: Nodes with more replies are considered more influential.
\item Distance from Root Node: Nodes closer to the original post have a higher influence on the overall emotional tone.
\item PageRank: A comment’s importance is determined by its position in the graph and the volume of engagement it receives.
\item Emotion Intensity: Each comment’s emotional intensity contributes to its influence score, affecting the root node’s overall emotion board.

This approach enables the model to identify specific nodes within the conversation likely to trigger emotional escalation or toxicity.
\end{itemize}

This graph-based structure allows us to assess how each comment influences the emotional tone of the conversation. These influence scores are used to update the emotion board of the root node, which compiles the emotional impact of all comments. To prevent the escalation of negative emotions, highly toxic comments are temporarily held before being added to the conversation graph, based on their impact on the root node.
\begin{figure}[htbp]
  
    \centering
    \includegraphics[width=13cm,height=13cm,keepaspectratio]{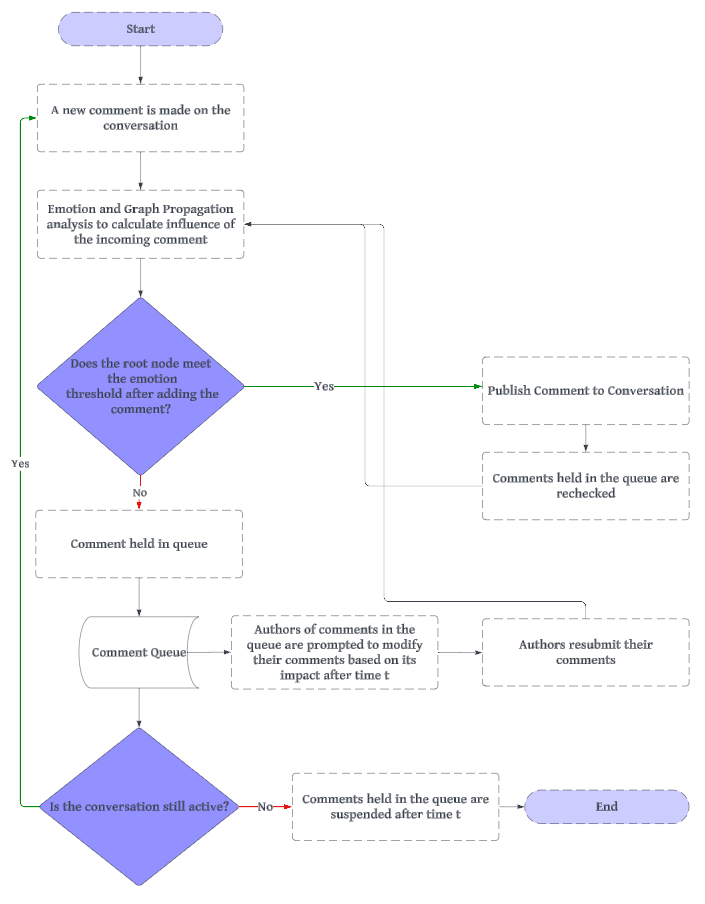}
  \caption{Proposed Comment Queuing to encourage Self-reflection}
  \label{fig:queue}
  \end{figure}
  
\subsection{Comment Queuing System for Self-Reflection}
To facilitate self-reflection and regulate negative emotions in online conversations, we introduce a comment queuing mechanism as shown in Fig. \ref{fig:queue}. This system assesses each new comment to calculate its potential impact on the overall emotional tone of the conversation before its publication. As each comment influences the emotional tone of the conversation, its inclusion also means an addition of emotional weight to the cumulative tone of the conversation. If the emotional impact of a comment pushes the predefined thresholds, such as Anger > 50\% or Fear > 60\%, the comment is flagged as toxic and temporarily stored in a queue. 

\begin{figure}[h!]
    \centering
    \subfigure[Angry comment queued]{
        \includegraphics[width=0.45\linewidth]{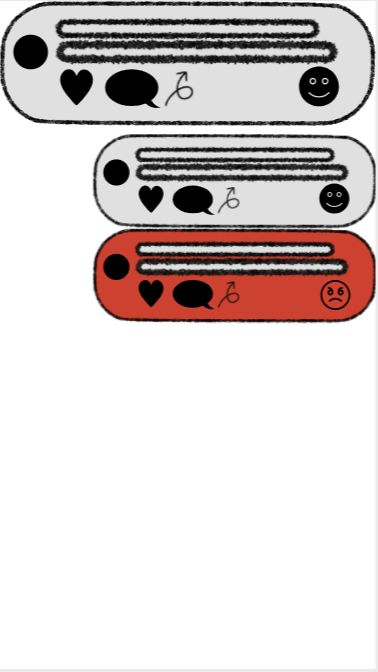}
        \label{fig:queued_comment}
    }
    \hfill
    \subfigure[Angry comment reintegrated]{
        \includegraphics[width=0.45\linewidth]{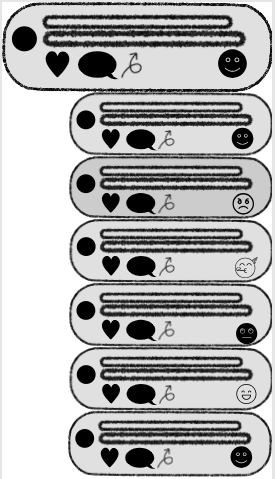}
        \label{fig:comment_reintegrated}
    }
    \caption{Comment queuing on a thread}
    \label{fig:side_by_side}
\end{figure}

For instance, as shown in Fig. \ref{fig:side_by_side}, in a Twitter conversation centred around government policy reforms, a reply that expressed strong anger toward a government policy raised the root node’s anger threshold to 65\%, Fig. \ref{fig:queued_comment}. As a result, it was queued, (indicated in red) This comment, positioned directly under the root node, significantly impacted the overall sentiment due to its emotional intensity. However, as can be seen in Fig. \ref{fig:comment_reintegrated}, when subsequent comments introduced feelings of trust and joy, the anger percentage fell below 55\%, enabling the queued comment to be integrated safely without further escalation.

While in the queue, comments are regularly re-evaluated each time a new comment is added to the conversation. This guarantees that the emotional board is continuously updated, enabling the system to determine whether previously queued comments can now be added without surpassing the thresholds. If the addition of a new comment balances or reduces the overall emotional intensity, the previously flagged comment can be safely integrated into the conversation.

To accommodate the flow of online conversations, the system distinguishes between active and non-active conversation stages. During active stages, the thresholds are more strict to prevent the escalation of negative emotions. As the conversation becomes less active, these thresholds may be relaxed, permitting more comments to enter the conversation while still upholding its emotional balance.

If a comment remains in the queue after all the other comments have been processed and still exceeds the emotional thresholds, its author is prompted to revise the comment. Following revision, the comment is re-evaluated. If the revised comment aligns with the emotional thresholds, it is included in the conversation; if not, the comment is suspended to prevent further emotional escalation and ensure a healthier dialogue.

The threshold values for the emotional tone of the conversation are determined using several parameters and are adjusted dynamically through an algorithm that takes into account the volume of comments, the general emotional distribution in the conversation, and recent variations in emotional intensity. For example, in very active discussions where numerous comments express high-intensity feelings, the thresholds for anger or fear might be temporarily increased to lessen the number of queued comments. On the other hand, during calmer times, the thresholds may be marginally reduced to enable stricter moderation and avert possible intensification of strong emotions. The active/non-active stage of the conversation is distinguished based on continuous analysis of time intervals between comments, overall comment volume, and user engagement patterns. This ensures the system can adapt to activity levels, such as high-intensity discussions or subdued phases. 

The threshold values also adjust to the conversation's context and evolving emotional distribution. A sliding window approach is employed, focusing on the most recent comments to capture real-time emotional states and prevent outdated emotions from distorting the tone. In our experiments, we used 100 recent comments. Positive emotions, like joy or love, are given higher weight through weighted allowances, promoting constructive interactions while moderating negative emotions like anger or fear. To minimise comment suspension and maintain emotional balance, the system prioritises underrepresented emotions in the conversation. This ensures that diverse emotional expressions are integrated, preventing the dominance of a single emotional tone while promoting a more balanced, inclusive dialogue.

By incorporating a reflective pause and offering users the opportunity to refine their contributions, this system promotes more deliberate and emotionally balanced interactions, while also discouraging impulsive comments.

For example, in a particularly active Twitter/Reddit thread discussing pandemic policies, the thresholds for anger and fear would be temporarily elevated due to the influx of comments expressing intense emotions. This adjustment will help reduce disruptions in the conversation, enabling comments with moderate emotional weight to be included while ensuring that extreme outliers to be managed and queued appropriately.

\subsection{Experimental Setup}
\begin{figure}[h]
  
    \centering
    \includegraphics[width=\textwidth,height=13cm,keepaspectratio]{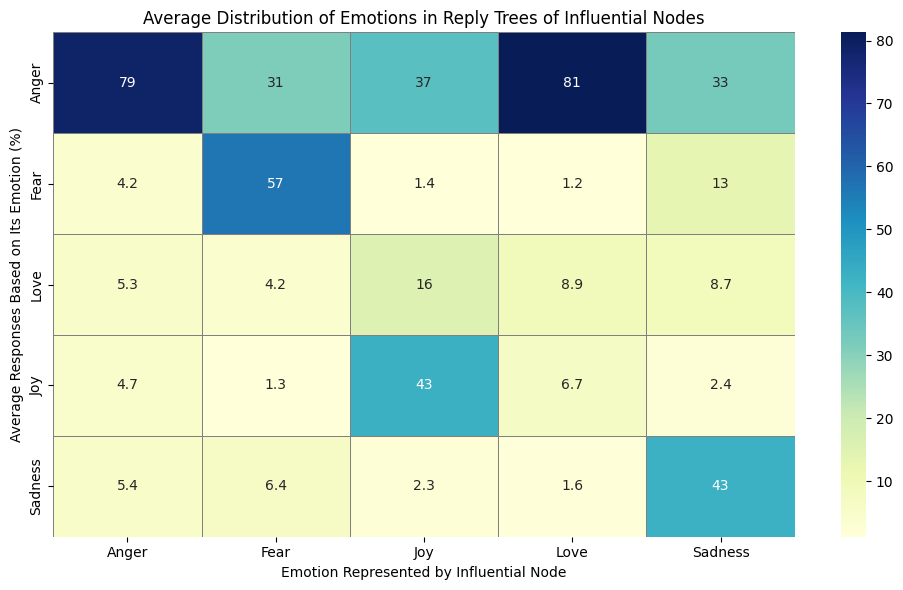}
  \caption{Average distribution of emotions and the number of unique users in the reply trees of influential nodes \cite{10520471}}
  \label{fig:heatmap}
  \end{figure}
As shown in Fig. \ref{fig:Framework}, following the processing of the collected data, each conversation was converted into a directed acyclic conversation graph. Subsequently, every comment in the graph was assigned an emotion using the NRC lexicon, along with the intensity of the emotion contained in the comment's content. The influence of each comment was computed based on a combination of its emotional intensity, distance from the root node, PageRank, and number of replies. This score was utilised to update the emotional impact of the conversation. The root node maintained an emotion board that tracked the percentage influence of each emotion on the conversation, with this board being dynamically updated as new comments were added to the graph. The framework was assessed by comparing its performance to the toxicity scores generated by the Google Perspective API \cite{bworld}. Our analysis focused on mitigating hate speech and polarisation by deactivating influential nodes or restricting responses once they reach the toxicity threshold. Our findings revealed that, while the Perspective API reduced hate speech by 7\%, the eImpact framework achieved a 10\% reduction. This confirms the effectiveness of the proposed framework in identifying and regulating post-toxicity while considering its subjectivity. Moreover, it can complement online content moderation efforts by providing insights into the sources of toxicity in conversations, taking both content and context into consideration. 

To simulate the evolving nature of online conversations, dynamic thresholds were introduced for detecting toxicity. These thresholds decreased as more comments were processed, enabling a flexible and adaptive assessment of emotional influence. In the scenario of queuing, comments that exceeded the toxic thresholds on the root node's emotion board were placed in a queue. These held comments were reassessed after each new comment addition, and those meeting the threshold were subsequently added to the graph.

\section{Results and Discussion}
In this study, we assessed the efficacy of the eImpact framework and the comment queuing mechanism in managing emotional dynamics within online conversations represented as graphs, with a specific emphasis on moderating toxic comments. We compared two different approaches: one without a queuing mechanism, where comments were added to the graph immediately, and one with a queue, where comments were temporarily held and reevaluated against emotional thresholds before being incorporated into the conversation.
\begin{table}[h]
\caption{Comparison of proposed framework \cite{10520471} against Perspective API \cite{bworld}}
\label{tab:my-table_comp}
\resizebox{\columnwidth}{!}{%
\begin{tabular}{llll}
\hline
Utilised Model &
  \begin{tabular}[c]{@{}l@{}}Identifying\\ influential nodes\end{tabular} &
  \begin{tabular}[c]{@{}l@{}}Toxic Node\\ Detection\\\end{tabular} &
  \begin{tabular}[c]{@{}l@{}}Estimated Toxicity\\ Reduction\end{tabular} \\ \hline
eImpact Framework&
  \begin{tabular}[c]{@{}l@{}}Based on\\ impact score,\\ taking into account\\ the tweet text \\ as well as its\\ connectivity\end{tabular} &
  1-4\% &
  10\% \\ \hline
Perspective API \cite{bworld} &
  \begin{tabular}[c]{@{}l@{}}Based on\\ toxicity, taking \\ into account only\\ the tweet text\end{tabular} &
  1-2\% &
  7\% \\ \hline
\end{tabular}%
}
\end{table}

Our analysis of emotional propagation in online conversations, using data from Twitter and Reddit, revealed that highly emotional content tends to receive greater user engagement. We observed that tweets featuring emotive language, such as expressions of love, often triggered intense, anger-fuelled reactions, leading to heightened emotional polarisation. This polarisation effect was pronounced when the same group of users repeatedly interacted with the post, resulting in the amplification of anger and the suppression of other emotions like joy. Our findings are consistent with existing research indicating that deeply rooted radical viewpoints are more likely to manifest as toxicity when anger dominates online discourse. Fig. \ref{fig:heatmap} shows the average distribution of emotions and the number of unique users involved in influential node reply trees for each emotion represented by the influential node.

  

\begin{figure}[ht]
    \centering
    \subfigure[Number of comments < 2000]{
        \includegraphics[width=0.45\linewidth]{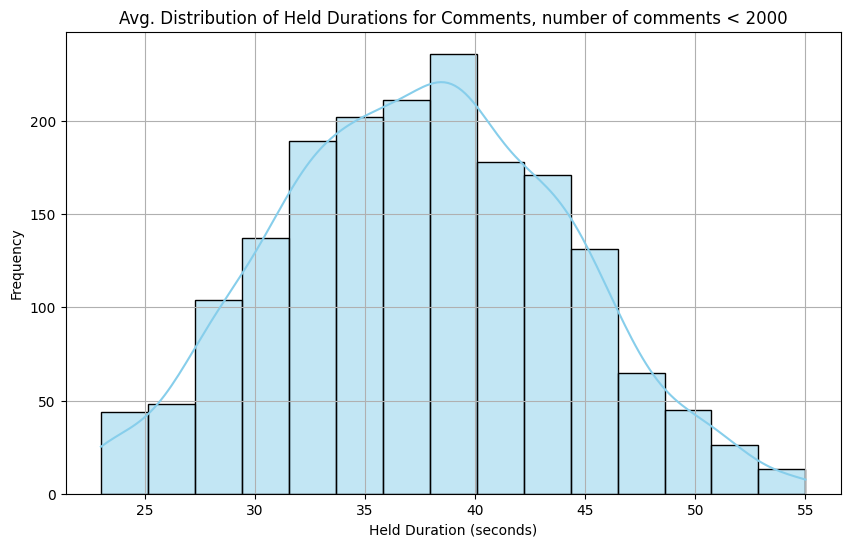}
    }
    \hfill
    \subfigure[Number of comments < 3000]{
        \includegraphics[width=0.45\linewidth]{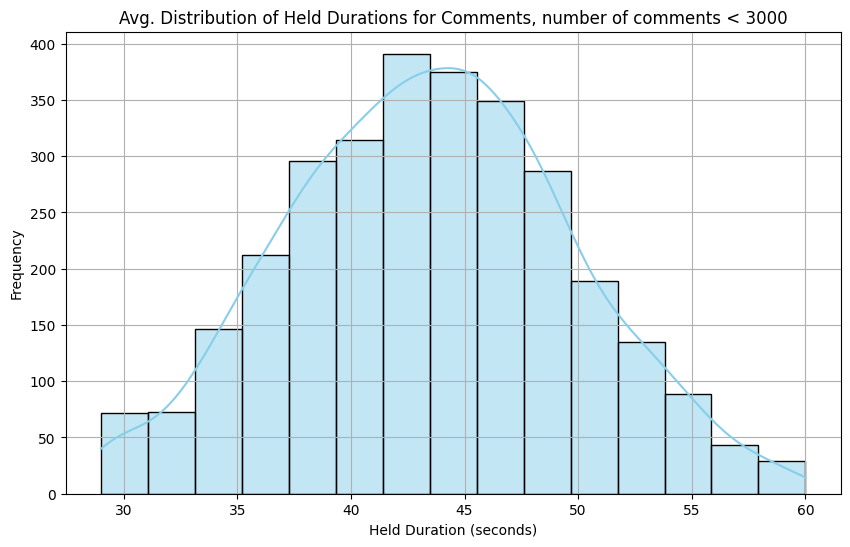}
    }
    
    \subfigure[Number of comments < 5000]{
        \includegraphics[width=0.45\linewidth]{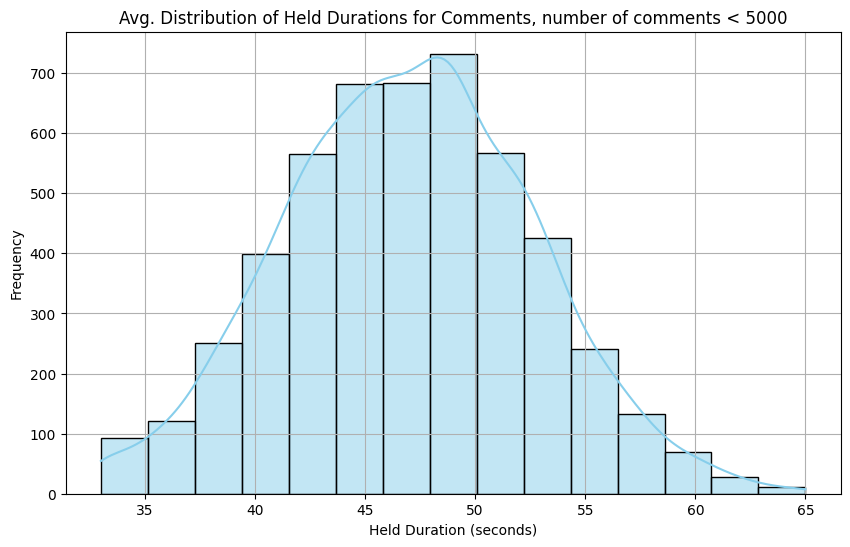}
    }
    \hfill
    \subfigure[Number of comments < 7000]{
        \includegraphics[width=0.45\linewidth]{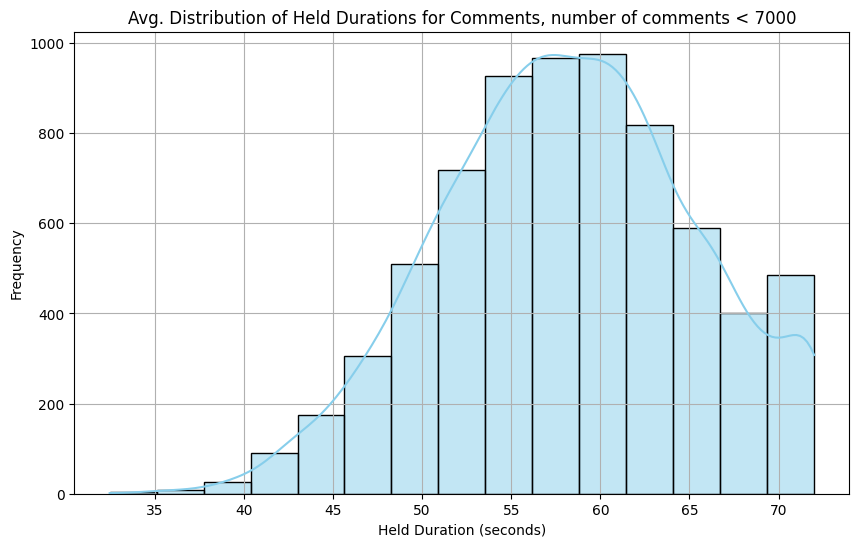}
    }
    
    \caption{Average Distribution of Held Durations for Comments when using the proposed queue approach}
    \label{fig:held_durations}
\end{figure}

We evaluated the eImpact framework, which is designed to mitigate emotional escalation, alongside Google's Perspective API for mitigating toxic content. The results demonstrated that eImpact led to a 10\% reduction in hate speech and polarising content, compared to a 7\% decrease with the Perspective API. This outcome emphasises the proposed framework's effectiveness in addressing not only the content but also the broader context and emotional dynamics within conversations. By prompting users to consider the emotional impact of their comments, eImpact promotes self-regulation and reduces toxic behaviour over time. Table \ref{tab:my-table_comp} compares the proposed framework to the Perspective API in terms of identifying influential nodes and the possibility of toxicity reduction.

\begin{figure}[h]
  
    \centering
    \includegraphics[width=12cm,height=12cm,keepaspectratio]{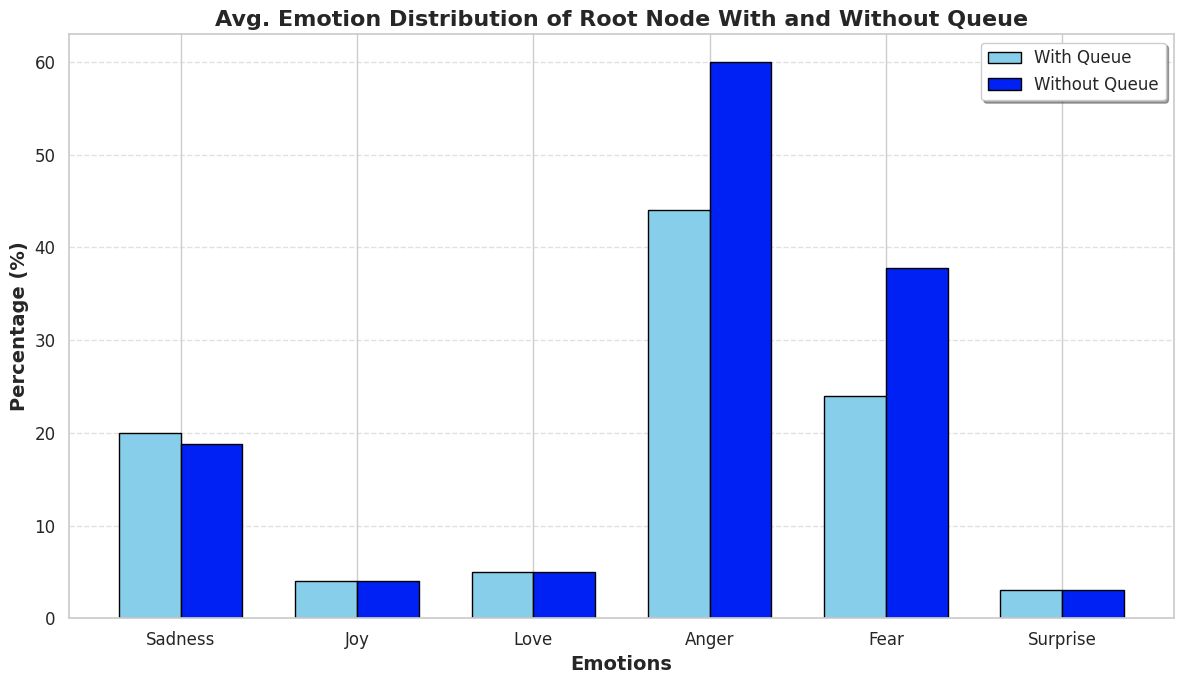}
  \caption{Average Emotion Distribution of Root Node With and Without Queue}
  \label{fig:emotion_board}
  \end{figure}

We then implemented the comment queuing mechanism to moderate emotional intensity in real-time. The queuing system managed comment posting by assessing their emotional impact before allowing them into the conversation. Analysis of the held durations revealed that comments were typically queued for 40 to 55 seconds, with an average hold time of 47 seconds, providing ample time for self-reflection. Fig. \ref{fig:held_durations} shows a histogram that illustrates the time distribution of comments held in the queue before being processed. The x-axis shows the duration of being held in seconds, while the y-axis represents the frequency of comments. As shown in the figure, significant concentration of comments with short-held durations indicates that the queue efficiently reintegrated comments into the flow. The distribution indicates that the majority of comments were held for brief periods, suggesting that the system effectively moderated emotions without causing substantial delays. Longer-held durations point to comments that posed greater emotional challenges, requiring extended holding times which is required to ensure the stability of the emotion board.

When comparing conversations with and without the queuing system, we observed that the emotion board maintained a more balanced distribution of emotions when the queue was employed. Dominant emotions such as anger and fear, which were prevalent without the queue, were significantly reduced. The gradual integration of comments into the conversation helped slow down emotional spikes, particularly in cases where comments could have rapidly escalated negative emotions. As a result, the conversation took on a more moderated tone. Fig. \ref{fig:emotion_board} showcases bar graphs that compare the final emotional composition of the conversation graph under the two experimental conditions—without queuing and with queuing. The y-axis shows the percentage contribution of each emotion to the overall emotional state of the root node. These graphs showcase the influence of the queuing mechanism on emotional dynamics. In the "Without Queue" scenario, the conversation is predominantly characterised by negative emotions such as Anger and Fear, while the "With Queue" scenario presents a more balanced emotional state. When the queue is used, emotions are always within threshold limits and not dominated by specific emotions. The findings suggest that in the absence of queuing, negative emotions tend to accumulate and dominate the conversation. Conversely, the queuing mechanism effectively moderates emotional impact, preventing negative emotions from overpowering the conversation and resulting in a more balanced emotional distribution. 

\begin{figure}[h]
  
    \centering
    \includegraphics[width=12cm,height=12cm,keepaspectratio]{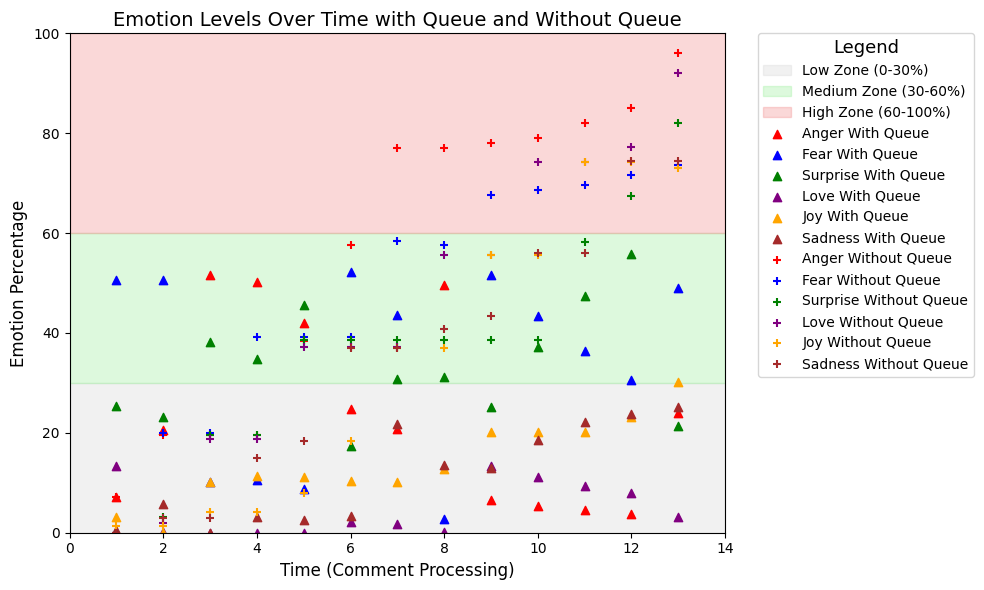}
  \caption{Average emotion levels in the conversation when using the queue}
  \label{fig:emotion_timeseries}
  \end{figure}

The dynamic thresholds, sliding windows, and weighted allowances in the queue mechanism contributed to sustaining the flow of the conversation. By adapting thresholds in response to the context and activity level of the conversation, the system was able to evaluate comments in real-time. The sliding window, which concentrated on the most recent 100 comments, ensured that only the most recent emotional trends influenced the conversation, thus preventing outdated emotions from influencing the overall tone. Moreover, by prioritising positive emotions and assigning higher priority to underrepresented emotions in the queue, the system facilitated an emotionally balanced conversation. Fig. \ref{fig:emotion_timeseries} shows a line graph illustrating the changing emotional impact over time for each emotion within the conversation graph when the queuing mechanism is implemented. The x-axis denotes time, while the y-axis represents the cumulative emotional impact. This graph offers a visualisation of how emotions evolve as the conversation unfolds and comments are added sequentially. It provides insight into how the queuing mechanism influences the flow of specific emotions. The graph indicates that negative emotions experience surges but are mitigated at various points, suggesting that the queuing mechanism intervenes to prevent them from dominating the conversation. On the other hand, positive emotions such as Joy and Love intensify over time, indicating that the system facilitates a more positive emotional trajectory. 

Overall, when the queue was employed, we observed an average reduction of 15\% in the spread of anger and fear compared to conversations where comments were posted immediately. This reduction highlights the potential of the queuing system to encourage healthier and less emotionally charged conversations by allowing users the time to reconsider their comments. Furthermore, only 4\% of comments were held for review, demonstrating the system's effectiveness in maintaining a smooth flow of conversation while managing emotional escalation. 

\subsection{Conclusion} 
This paper presents an innovative framework that combines comment queuing with adaptive thresholds to decrease online toxicity and facilitate emotion regulation in digital conversations. By holding potentially harmful comments in a queue and evaluating their impact before publishing, the framework promotes self-reflection and mitigates emotional spikes. Our experiments showed a 15\% reduction in negative emotions such as anger and fear compared to unmoderated conversations.

This framework makes a valuable contribution to the field of Digital Emotion Regulation (DER) by encouraging healthier and more responsible online discourse without compromising user engagement. Future research can build upon this work by implementing adaptive mechanisms for different types of content and exploring real-time emotional feedback through user interviews and surveys.

\section{Acknowledgement}
During the drafting of this paper, Grammarly \cite{grammarly} was used to check and enhance the grammar and writing style of this document. We are grateful to Amity University for accepting our work as a keynote paper.
%
%
%
\bibliographystyle{splncs04}
\bibliography{refs.bib}

\end{document}